\documentclass[amsmath,amssymb,aps,prb,floats,floatfix,twocolumn,superscriptaddress]{revtex4-2}

\usepackage{graphicx}
\usepackage{bbm}
\usepackage[utf8]{inputenc}
\usepackage[T2A,T1]{fontenc} 
\usepackage{amssymb}
\usepackage{amsmath}
\usepackage{graphics}
\usepackage{times}
\usepackage{bbm}
\usepackage{tabularx} 
\usepackage{tabu}
\usepackage{multirow}

\usepackage{hhline}
\usepackage{psfrag} 
\usepackage{eurosym}
\usepackage{titlesec}
\usepackage{titletoc}
\usepackage{tocloft}
\usepackage{adjustbox}
\def\€{\euro{}}
\usepackage{url}
\usepackage{color}
\usepackage[dvipsnames]{xcolor}
\usepackage{hyperref}
\usepackage[normalem]{ulem}
\usepackage{float}
\definecolor{linkcolor}{rgb}{0,0,0.6} % définition de la couleur des liens pdf
\definecolor{WhitishBckgnd}{rgb}{0.999,0.999,0.999}  
\usepackage{fancyhdr} %en-tête de page personnalisés 
\hypersetup{ 
    colorlinks, 
    citecolor=black, 
    filecolor=black, 
    linkcolor=black, 
    urlcolor=linkcolor 
}   
\DeclareGraphicsExtensions{.jpg, .eps}
\usepackage{siunitx}
\usepackage{booktabs}% http://ctan.org/pkg/booktabs

\renewcommand{\nabla}{
\hspace{-1ex}\stackrel{\mbox{\raisebox{-1.0ex}[-2.5ex][0ex]{\del}}}{\mbox{\raisebox{0.45ex}{\dell}}}\hspace{-1ex}
}

\newcommand{\cogd}{Co$_{1-x}$Gd$_x$~}

\begin{document}
%\renewcommand{\thepage}{\roman{page}}

%\renewcommand{\cftsecleader}{\cftdotfill{\cftdotsep}}
%\renewcommand{\cftdot}{.} %empty {} for no dots. you can have any symbol inside. For example put {\ensuremath{\ast}} and see what happens.
%
%
%\title{Publication outline - XRMS-PEEM on CoGd beads}

\title{Small-angle X-ray resonant magnetic scattering at the Co M$_{2,3}$ and L$_{3}$ edges observed with photoemission electron microscopy}

\author{Alexis Wartelle}\altaffiliation[Present address: ]{Université Grenoble Alpes, CNRS, Grenoble INP, Institut Néel, Grenoble, France}
\affiliation{Institut Jean Lamour (UMR CNRS 7198), Université Lorraine, 54000 Nancy, France}
\affiliation{European Synchrotron Radiation Facility, F-38043 Grenoble, France}
\email{alexis.wartelle@ens-lyon.org}
\author{Marisel Di Pietro Mart\'{i}nez}
\affiliation{Max Planck Institute for Chemical Physics of Solids, 01187 Dresden, Germany}
\author{Olivier Fruchart}
\affiliation{Univ. Grenoble Alpes, CNRS, CEA, Spintec, Grenoble, France}
\author{Philippe David}
\affiliation{Université Grenoble Alpes, CNRS, Grenoble INP, Institut Néel, Grenoble, France}
\author{Guillaume Beutier}
\affiliation{Univ. Grenoble Alpes, CNRS, Grenoble INP, SIMaP, 38000 Grenoble, France}

% XMCD: X-ray magnetic circular dichroism
% X-ray magnetic singular linear polarization ptychograpy with mode analysis.
% single-polarization X-ray ptychograpy
% imaging the magnetization with single linear polarization ptychograpy by using a multimode analysis

\date{\today}% It is always \today, today,
             % but any date may be explicitly specified

\begin{abstract}
%%% Try not to go more than 3500 words.
X-ray magnetic circular dichroism is an efficient contrast mechanism allowing for a direct sensitivity to magnetization. Combined with an imaging technique such as photoemission electron microscopy, it has been successfully applied to high-resolution investigations of ferromagnetic thin films but also of three-dimensional systems thanks to the transmission-type contrast in their shadow. Our focus in this work is the wave-optics scattering pattern that can be observed near such a shadow's rim. Taking advantage of non-uniform magnetic states present in near-micron-size Co$_{1-x}$Gd$_x$ beads, we first show how X-ray resonant magnetic scattering affects the Fresnel diffraction at the Co L$_3$ edge. In order to confirm this observation, we then turn to the Co $M_{2,3}$ edges. There, we measure magnetic scattering patterns with a significantly increased spatial extent (due to the larger wavelength), despite the signal's weakness. The patterns' origin is supported by a comparison between our experimental data and a simple analytical model, then numerical simulations.
\end{abstract}

\maketitle

\section{Introduction}

Among the properties of matter which may be probed with X-rays, magnetism holds a somewhat peculiar place. In many, if not most situations, the interactions between such radiation and magnetic moments are several orders of magnitude weaker than the interactions originating from the material's electronic density~\cite{Als-Nielsen2011}. Yet, a variety of approaches have been developed over the years to explore magnetic orders using X-rays, from spectroscopy~\cite{Baudelet1991_XMCDatNdL2Edge,Chen1995} to diffraction~\cite{Johnson2011_NonResonantMagDiffraction,Biffin2014_ResMagDiffraction_MagOrderInBeta_Li2IrO3,Beutier2017_XrayDiffraction_weakFMs,Misawa2021_CuL3edge_ResonantDiffraction} through imaging~\cite{Kimling2011,Streubel2015a,Hierro-Rodriguez2017,Donnelly2017}. The latter field exploits the chemical selectivity and strongly enhanced scattering cross-sections~\cite{Stoehr2006,van_der_Laan2008_XRMS_formalism_review_Chirality} available at photon energies close to certain absorption edges of magnetic elements, where X-ray Resonant Magnetic Scattering (XRMS) occurs. Of particular interest are the $L_{2,3}$ absorption edges of 3d transition metals such as Fe, Co, and Ni, as well as the $M_{4,5}$ edges of rare-earth elements, such as Gd, Dy or Ho. These have been of crucial importance in the development of modern X-ray magnetic microscopy techniques~\cite{Cheng2012,Donnelly2019_ReviewOn3DxRayMagImaging_Tomography}, which use XRMS as a contrast mechanism. 

The most frequently used manifestation of XRMS is X-ray Magnetic Circular Dichroism (XMCD)~\cite{Stoehr2006}, which is well-suited to several X-ray microscopy techniques \emph{e.g.} (Scanning) Transmission X-ray Microscopy [(S)TXM] or PhotoEmission Electron Microscopy (PEEM)~\cite{Donnelly2019_ReviewOn3DxRayMagImaging_Tomography}. However, the phase contrast counterpart of XMCD, namely X-ray Magnetic Circular Birefringence (XMCB), may also be used in experiments sensitive to sample refraction. This is relevant notably for Fourier Transform Holography (FTH)~\cite{Stroke1964_FTHprinciple,GuizarSicairos2007_HERALDO} and ptychography~\cite{Rodenburg2007_PtychographyOnExtendedObjectsWithHardXrays,Donnelly2019_ReviewOn3DxRayMagImaging_Tomography}. As was demonstrated both in magnetic FTH~\cite{Scherz2007_FTH_Co,DiPietroMartinez2023_3D_FTH_onFeGdML} and in ptychography~\cite{Donnelly2016,Neethirajan2023}, XMCB may not only yield substantial magnetic contrast, but also allow to work at photon energies significantly below the relevant peak of X-ray absorption. This in turn enables transmission imaging of relatively thick samples~\cite{DiPietroMartinez2023_3D_FTH_onFeGdML,Neethirajan2023}.

While the typical spatial resolutions available with magnetic X-ray microscopy are not better than in electron-based techniques like electron holography and Lorentz microscopy~\cite{Hubert1998} or even (depending on the case) in Magnetic Force Microscopy (MFM)~\cite{Abelmann2005}, their primary advantage lies with the direct sensitivity to magnetization, as opposed to induction along the electron path or to magnetostatic charges. Furthermore, as the X-ray attenuation lengths vary (for absorption edges relevant for XRMS) from \emph{ca.} a dozen nanometres to several hundreds of nanometres in the soft X-ray regime, there is ample freedom to adjust the probed depth of material experimentally.

In the specific case of X-ray PEEM, this large variability provides a unique versatility when three-dimensional samples are studied. Indeed, XMCD-PEEM can usually yield a (mostly surface-related) image of the sample viewed from the top, but the transmission-type magnetic information in the object's elongated shadow may or may not be retrieved depending on signal-over-noise ratio. If the sample's absorption is too strong, there is no shadow contrast~\cite{Hertel2005_XMCD_PEEM_FeDots_NoShadowContrast}, whereas larger X-ray transmission allows for valuable, volume-related magnetic signal~\cite{Kimling2011,DaCol2014,Bran2016,Wartelle2019}. However, strictly speaking, shadow XMCD-PEEM images are not simply an elongated transmission view of a 3D object. As was noted by Jamet \emph{et al.}~\cite{Jamet2015}, sub-micron objects do create Fresnel diffraction patterns in the close vicinity of their shadow's rim. There appeared to be magnetic scattering in the corresponding XMCD-PEEM images at the Fe L$_3$ edge, however the confinement of the interference pattern is an obstacle to investigating small-angle XRMS from 3D objects with PEEM.

Considering the above, it is rather natural to look for a possibility to (i) decrease the Fresnel number~\footnote{We take here the following definition for the Fresnel number $F$: $F=d^2/(4b\lambda)$~\cite{handbook_of_optics_vol1} where $d$ is the typical object size, $b$ the distance from it to the plane of observation, and $\lambda$ is the relevant wavelength.  With the $\Psi=$\ang{16} incidence angle in PEEM, $F\simeq 6.9\cdot d/\lambda$ when considering the shadow's rim.} to suppress this confinement (ii) while preserving magnetic sensitivity (iii) as well as coherence. One route, when considering 3d transition metals, consists in moving from their L$_{2,3}$ edges down to M$_{2,3}$ edge. Indeed, the wavelength increases by an order of magnitude, XRMS is still sizeable~\cite{Valencia2006_Medge_XMCD_MagnetoOpticalConstantsOfFeNiCo}, and the coherence of synchrotron radiation is only expected to increase significantly~\cite{Als-Nielsen2011}.

We point out that the relevance of XRMS at the Co M$_{2,3}$ edges for imaging has already been demonstrated by the work of Kfir \emph{et al.}~\cite{Kfir2017_HHG_FTH_CoPd}, who employed extreme-ultraviolet FTH to recover the domain pattern in a Co/Pd multilayer. It was shown~\cite{Kfir2017_HHG_FTH_CoPd} that the application of Coherent Diffraction Imaging (CDI~\cite{Chapman2006_AbInitioXrayCDI}) algorithms to the collected diffraction data (with the FTH images as starting point) enabled substantial increase both in signal-over noise ratio and spatial resolution. With respect to this approach and similar reports~\cite{Tadesse2018_XUV_FTHwithCDI,Eschen2021_TR_XUV_FTH} utilizing high-order harmonic generation~\cite{McPherson1987_HHG}, synchrotron-based PEEM at the same photon energy is expected to deliver a three orders of magnitude better energy resolution and a much larger photon flux. Last but not least, phase retrieval from small-angle scattering patterns acquired in a reflection geometry has been demonstrated by Mente\c{s} \emph{et al.}~\cite{Mentes2020_XRDinSPELEEMonArtifSpinIces}. There, the authors needed PEEM images of their artificial spin ice systems as input for the CDI algorithms. However, for a given sample-to-detector distance, the oversampling criterion~\cite{Chapman2006_AbInitioXrayCDI} is harder to fulfil at the Fe L$_3$ edge than at its M edge. Therefore, it seems that CDI in the above-mentioned, PEEM-based reflection geometry may become relevant for magnetic imaging at such energies. Two strong arguments to pursue this are (i) the fact that this modus operandi allows imaging under field since this is a photon-in, photon-out technique~\cite{Mentes2020_XRDinSPELEEMonArtifSpinIces} as opposed to regular XMCD-PEEM, and (ii) the absence of need for alignment of many sample images before computing a difference image (to reveal magnetic contrast) with good stastistics.

With the above in mind, we explore in this work X-ray Resonant Magnetic Scattering (XRMS) from 3D objects at the Co M$_{2,3}$ edges, using a PEEM for imaging. To the best of our knowledge, this the first experimental application of extreme UV light to magnetic Fresnel diffraction in the specific configuration imposed by X-ray PEEM. The manuscript is organized as follows. To begin with, we describe the Co$_{1-x}$Gd$_x$ bead samples which are used in this work, as well as the lab- and synchrotron-based imaging methods that have been used. We then move on to the results that have been obtained, starting with a pre-characterization, then presenting the magnetic PEEM imaging. In this respect, we establish a solid overview of our samples' magnetic states with XMCD-PEEM at the Co L$_3$ edge, and as a second step perform imaging at the Co $M_{2,3}$ edges. There, we  present our interpretation of the obtained scattering patterns.

\section{Methods}

\subsection{Samples}

Our sample fabrication began with Si substrates onto which a Ta\textbackslash Pt bilayer was deposited for the sake of surface electrical conductivity. Based on prior calibrations of deposition rates in the used magnetron sputtering, the nominal stacking should be Ta(\SI{6}{\nano\meter})\textbackslash Pt(\SI{9}{\nano\meter}). Then, a pattern consisting of alignment marks was etched into the bilayer until the underlying Si became exposed, thanks to optical lithography and subsequent ion beam etching. During this second step, mass spectrometry allowed to interrupt the etching when all of the Ta had been removed inside the resist's patterned grooves.

Substrate pieces of adequate size were then introduced into a Pulsed Laser Deposition (PLD) system. There, large-fluence laser shots on a target with composition Co$_{80}$Gd$_{20}$ led to the formation of Co$_{1-x}$Gd$_x$ beads on the substrate. Although we expect their stoichiometry to not differ too much from the target's, we do not have quantitative composition measurements on these samples~\footnote{Independent X-ray ptychographic imaging performed on one such \cogd bead at the Gd L$_3$ edge (not shown here) suggests a homogeneous composition in the volume.}. After several dozens of laser impacts on this first target, a Nb coating of a few nanometres was deposited with PLD so as to obtain a passivation layer and partially protect the beads from oxidation. 

\begin{figure}[h!]
\centering\includegraphics[width=\columnwidth]{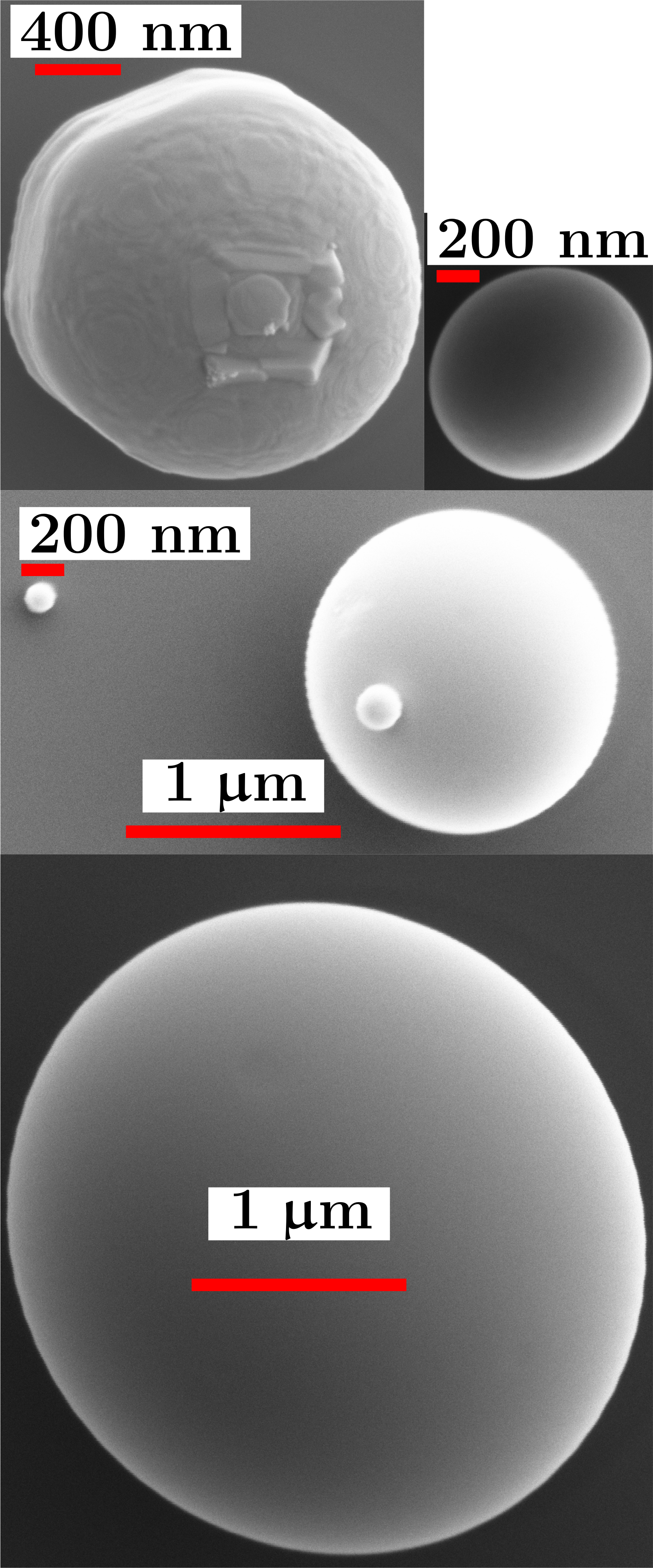}
\caption{Scanning electron micrographs of beads from PLD. All these images are on the same scale}\label{fig_SEM_beads}
\end{figure}

The PLD approach leads to a wide variety in terms of shapes and sizes. A small glimpse into the latter is provided in Fig.~\ref{fig_SEM_beads}, whose Scanning Electron Microscopy (SEM) images display objects with lateral dimensions ranging from \SI{200}{\nano\meter} to \SI{3}{\micro\meter}. Since these images do not convey a full sense of shapes, we add that the beads may be close to spherical, flattened and globular, or (much) more distorted. Furthermore, a small proportion of them is somewhat polyhedral in shape, as is exemplified by the bead on the top left of Fig.~\ref{fig_SEM_beads}. Though interesting in its own right, this stark contrast with the smoother and more round majority of objects goes beyond the scope of this manuscript. We point out that this collection is representative of objects on which Energy-Dispersive X-ray analysis was performed and confirmed the presence of \cogd ; however the specific beads shown in Fig.~\ref{fig_SEM_beads} have not been checked in this fashion themselves. There is a possibility that some of these are actually pure Nb, as we have determined that such beads may also be present. Yet, we insist that observations of dozens of \cogd beads lead us to consider these SEM image as typical.

\subsection{Magnetic Force Microscopy imaging}

Due to the aforementioned diversity in \cogd bead shapes and sizes, it would not be very informative to try to perform magnetometry measurements \emph{e.g.} with a Vibrating Sample Magnetometer. Since magneto-optical measurements would be hindered by the curved bead surfaces as well as the typical size being at most a couple of times the light's wavelength, we turned to Magnetic Force Microscopy (MFM). Our goal was gain some insight into the magnetic configurations that can be present in our samples.

All of our MFM acquisitions followed the common two-pass scheme. The Atomic Force Microscopy (AFM) and MFM images presented in Section \ref{sec_results} were obtained with a custom-coated probe. A \SI{25}{\nano\meter}-thin Co$_{80}$Cr$_{20}$ layer was sputter-deposited on an AC240TS AFM tip from Asylum Research before the addition of a few nanometres of SiO$_2$ for protection against oxidation. The direction of the tip's magnetization was controlled by setting it in close proximity to a permanent magnet (with unambiguously identified poles) before imaging. The tip's oscillation amplitude was \SI{25}{\nano\meter}, the second-pass tip lift was set to \SI{11}{\nano\meter}, and the recorded MFM signal was the tip's oscillation phase with respect to the phase of its AC excitation. Therefore, a negative phase corresponds in our convention to attractive forces~\cite{Thiaville2005_MFMchapter}. We point out that the MFM data presented in this manuscript was acquired after removing the electrostatic contribution to the oscillation phase thanks to a suitably chosen applied bias voltage between tip and sample~\cite{Jaafar2011_KelvinProbeForceMicroscopy_CancellingElectrostaticsForMFMonNonFlatSamples}.

In addition to this dataset, MFM images of \cogd beads acquired with a different instrument are presented in the Supplemental Material. 

\subsection{Synchrotron-based imaging and spectroscopy}\label{sec_Methods_PEEM_imaging}

All the synchrotron experiments relevant to this work were performed at the undulator beamline
Nanospectroscopy of Elettra (Sincrotrone Trieste). Our first step consisted in identifying \cogd beads with certainty; to that end, we performed X-ray Absorption Spectroscopy around the Co L$_3$ edge. Based on the obtained absorption spectra recorded on the top of the beads, we performed XMCD-PEEM imaging at energies up to the absorption peak, and as far down as \SI{3.4}{\electronvolt} below that peak. The reason for this is the sharply-varying X-ray penetration depth in this photon energy range. To set the scales, let us recall that for pure Co, taking magnetism into account, the X-ray attenuation length varies from \SI{0.51}{\micro\meter} at the foot of the L$_3$ edge to as low as \SIrange{12}{21}{\nano\meter} at the absorption peak ~\cite{Nakajima1999} \emph{i.e.} over less than \SI{6}{\electronvolt}. 

As a result, we go from sensitivity to only the sample volume (in shadow XMCD-PEEM) well below the absorption peak to pure surface sensitivity at the top of the L$_3$ edge's white line. We note that the most suitable shift in photon energy (with respect to the peak of absorption) for shadow XMCD-PEEM imaging depends on both the bead's geometry and its composition. In any case, this optimization of magnetic contrast as a function of energy allows to probe relatively large depths of material, as has been reported by several groups~\cite{DiPietroMartinez2023_3D_FTH_onFeGdML,Neethirajan2023}. For each acquisition, tens of individual snapshots with exposure times on the order of a second were first recorded, the first half with a given sign of photon helicity, the second one with opposite photon helicity. The polarization state available on the Nanospectroscopy beamline is elliptical, with a degree of circular polarization of about 0.77 around the Co L$_3$ edge. In order to compensate for sample drift during exposure, each set of images was aligned using sharp features such as bead edges as references. Averaging the aligned series resulted in two micrographs $I_+$ and $I_-$ corresponding to opposite helicities; the latter two were then aligned before the asymmetry ratio $\mathcal{A}=(I_+-I_-)/(I_++I_-)$ was computed for each pixel. In the following, we will refer to $\mathcal{A}$ as ``magnetic image''. The methodology for the final alignment consisted in minimizing the typical antisymmetric contrast in $\mathcal{A}$ that arises from under-/overcompensation of drift in both the horizontal and vertical directions. 

Due to the strong absorption of Nb in the vicinity of the Co M$_{2,3}$ edges (at about \SI{60}{\electronvolt}), we were initially not able to detect the latter with XAS. Therefore, we first performed three \SI{30}{\minute}-long rounds of etching via Ar plasma, at operating pressures $\sim$\SI{3}{\milli\bar} and accelerating voltage \SI{1.5}{\kilo\volt}, and a final  \SI{30}{\minute} round at an accelerating voltage of \SI{1.2}{\kilo\volt}. Thereafter, the observation of the Co M$_{2,3}$ edges indicated that magnetic imaging at these energies had become possible.

The Co M-edge magnetic images were acquired and processed similarly to the L$_3$ edge data. However, the total exposure times were twice larger for the sake of signal-over-noise ratio. Concerning image processing, we point out that the final alignment of M-edge data was determined not only by considering the above-mentioned criterion regarding drift-related asymmetric contrast but also the behaviour of $\mathcal{A}$ on the substrate, away from beads, their shadow or interference patterns. For a set $\mathcal{P}$ of pixels corresponding to a (reasonably small) patch on the substrate, if the sample drift is nearly compensated, then we may expect the following behaviour for the summed values of $\mathcal{A}^2$ over $\mathcal{P}$:

\begin{equation}\label{eq_SumXMCDsq}
\sum_{(x_i, y_i)\in\mathcal{P}}\mathcal{A}^2(x_i, y_i)\simeq \mathcal{A}_0^2+c_X\cdot(X_\mathrm{d}-X_0)^2+c_Y\cdot(Y_\mathrm{d}-Y_0)^2
\end{equation} \\ where $c_X$ and $c_Y$ are positive constants that are related to the substrate's roughness, and $(X_\mathrm{d}-X_0)$ and $(Y_\mathrm{d}-Y_0)$ are respectively the difference (in the horizontal/vertical direction) between the shift applied to align the $I_+$ and $I_-$ images and the optimal shift to fully compensate the sample drift. Finally, $\mathcal{A}_0$ is a background level that is not necessarily null. Indeed, any slight imbalance in photon helicity and any change in illumination profile may lead to a small background. However, whatever this (non-physical) contrast value, minimizing the left-hand side in Eq.~\eqref{eq_SumXMCDsq} for a given acquisition should suppress any artificial contrast. 

We stress that this refinement becomes necessary for M-edge data for several reasons. First of all, the small attenuation length at these photon energies prohibited any shadow XMCD-PEEM: in these magnetic images, we sought XRMS contrast in the interference pattern behind a bead's shadow, corresponding to much lower signal-over-noise ratio than for L$_3$ edge (shadow) XMCD-PEEM. In addition, we encountered instrumental difficulties during acquisition that were due to the strongly enhanced beamline flux (and the resulting larger current in the imaging column) in the extreme ultra-violet range~\cite{Locatelli2011_ChargingEffectsInPEEMdueToLargeFlux}. With respect to this report by Locatelli \emph{et al.}, we expect the incident flux to be comparable in our experiments, and indeed we found a degraded image quality at the Co M$_{2,3}$ edges that could not be improved upon through the objective lens. Moreover, we experienced a lower-than-usual positional stability of the microscope's angle-limiting aperture (also known as contrast aperture), which we ascribe to the greatly-enhanced electron currents in these conditions. It must be noted that this aperture is of crucial importance in our magnetic imaging experiments on 3D objects since we rely on sample edges as well as substrate roughness (\emph{cf.} previous paragraph) to refine the image alignment, and the aspect of such topographic features in PEEM images strongly depends on this aperture~\cite{Schneider2002_MagneticImagingAndResolutionInXPEEM,Jamet2015}.

\section{Results and discussion}\label{sec_results}

\subsection{Magnetic Force Microscopy}

\begin{figure}[!h]
\includegraphics[width=\columnwidth]{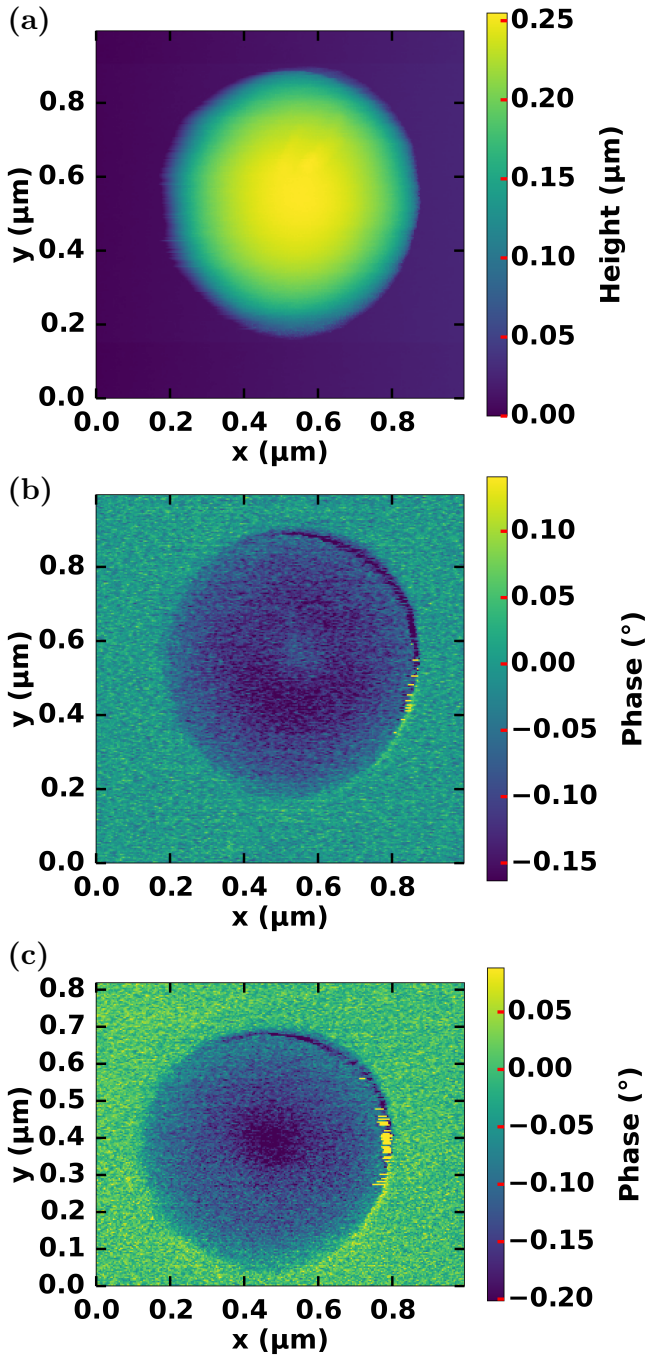}
\caption{(a) AFM image on a \cogd bead. (b) MFM image corresponding to (a) and acquired with an upward-magnetized tip.  (c) MFM image of the same bead acquired with a downward-magnetized tip.}\label{fig_MFM}
\end{figure}

Several \cogd beads that have been imaged with MFM yield similar maps, featuring a background all over the object (which to a good approximation, appears proportional to the local bead height) and a clear central bump. To illustrate this, we present in Fig.~\ref{fig_MFM}.(a) an AFM image of a bead on which MFM has been performed twice, with opposite tip magnetizations. The corresponding magnetic images in Fig.~\ref{fig_MFM}.(b-c) clearly show that the background's sign is independent on the tip's magnetization, whereas the central bump's is not. Therefore, the former can be ascribed to a susceptibility contribution, which is always attractive and is indeed negative~\cite{Thiaville2005_MFMchapter}, while the latter unambiguously indicates stray fields from the bead's magnetization. At this stage, bearing the similarity to published MFM images made on vortices~\cite{Shinjo2000_MFM_magneticVortices,Skidmore2004_MFM_NiDotsTargetDomains} as well as the bead's platelet shape in mind, it seems quite likely that this sample hosts a magnetic configuration comprising a vortex pointing perpendicular to the substrate.

We note that the lack of quantitative knowledge regarding this specific bead's composition and magnetic properties prevents us from backing this interpretation with direct arguments. However, we point out that the vortex state has been shown to persist in a wide variety of geometries and even in the presence of bulk and/or surface anisotropies ~\cite{Streubel2012,Pylypovskyi2015,Berganza2022_SkyrmionLikeConfigsInPyDomes,Finizio2022} at the cost of more or less pronounced distortions. Thus, even if the \cogd alloy is not a magnetically soft material, it does not necessarily prevent the existence of a vortex or vortex-like configuration. Finally, such a state has the advantage of significant flux-closure (which is compatible with the absence of magnetostatic charges over most of the bead) and it is the simplest pattern that is compatible with our MFM images. As a matter of fact, departures from the textbook vortex~\cite{Hubert1998} are not important to us, since we only need non-uniform magnetic configurations to look for XRMS in PEEM.

\subsection{Spectroscopy}\label{sec_Results_Spectro}

\begin{figure}[!h]
\includegraphics[width=\columnwidth]{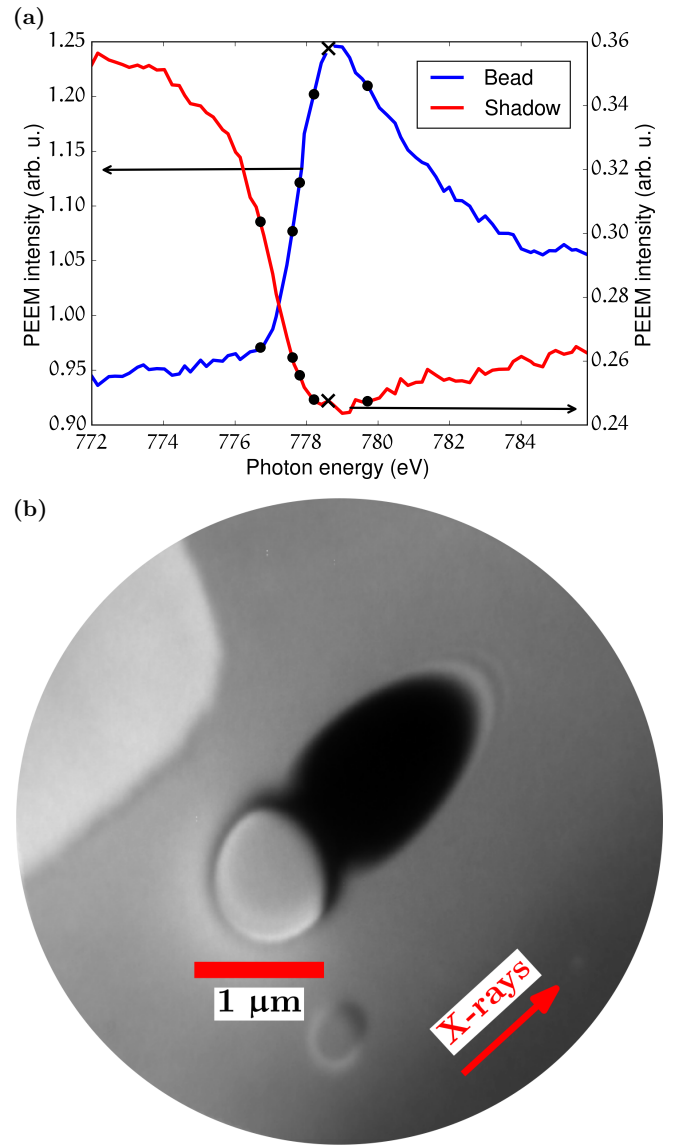}
\caption{(a) Typical XAS traces obtained respectively from the top of a \cogd bead (blue) and from its shadow (red). Full black dots $\bullet$ highlight photon energies at which XMCD-PEEM was performed (\emph{cf.} Fig.~\ref{fig_XMCD_PEEM_L3}), while a cross $\times$ indicates the photon energy \SI{778.6}{\electronvolt} at which the image from Fig.~\ref{fig_XAS_PEEM_bead}.(b) was acquired. (b) PEEM image at Co L$_3$ edge's absorption peak of the bead (in the center) from which the XAS data from from Fig.~\ref{fig_XAS_PEEM_bead}.(a) was extracted. The adjacent wide patch on the top left of the bead corresponds to an alignment mark, and the smaller round item below the bead and slightly to the right is another \cogd object.}\label{fig_XAS_PEEM_bead}
\end{figure}

The first result from our L$_3$ XAS measurements is the absence of Co oxidation near the beads' surface in most (if not all) cases. This is assessed from the peak shape of the photoemission recorded from the top of the objects: we usually do not observe the easily-recognizable structure that is characteristic of cobalt oxide~\cite{Regan2001}. This confirms the efficiency of our Nb capping. As an example, Fig.~\ref{fig_XAS_PEEM_bead}.(a) provides direct and transmission XAS traces obtained on the \cogd bead shown in Fig.~\ref{fig_XAS_PEEM_bead}.(b). As expected, the shadow spectrum saturates because of the bead's size [\SI{0.9\pm 0.1}{\micro\meter}]. However, the direct XAS is a good indicator of the absorption maximum, and does not show any satellite peaks. When imaged at the edge's peak [highlighted with a black cross in Fig.~\ref{fig_XAS_PEEM_bead}.(a)], our \cogd beads display a low-intensity shadow with weak fringes close to their rim, as visible in Fig.~\ref{fig_XAS_PEEM_bead}.(b).

\begin{figure}[!h]
\includegraphics[width=\columnwidth]{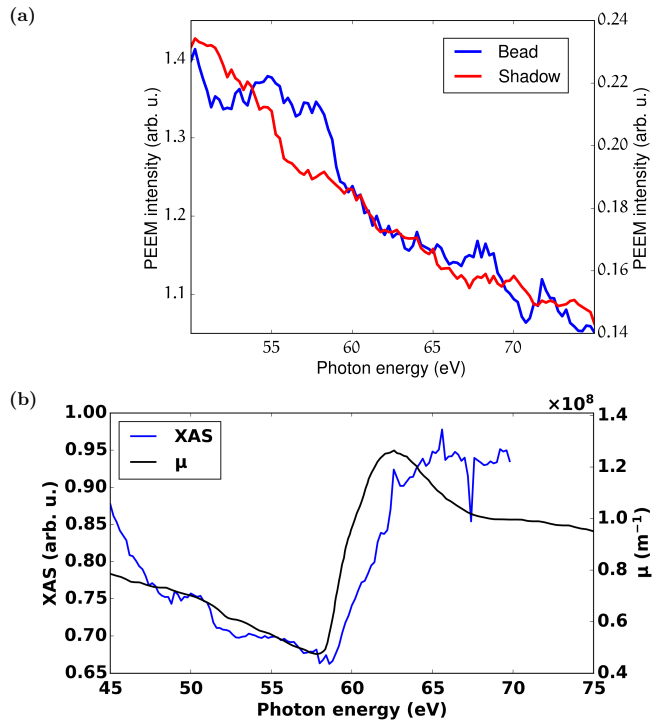}
\caption{(a) XAS traces around the Co M$_{2,3}$ edges obtained on a \cogd bead before sputtering. It must be noted that several spectra qualitatively similar to this one were obtained on different beads. (b) XAS trace (blue curve) around the Co M$_{2,3}$ edges obtained on a different \cogd bead after several rounds of Ar plasma sputtering. The significant (in terms of relative change) upturn in PEEM intensity starting a bit before \SI{60}{\electronvolt} is attributed to the Co M$_{2,3}$ edges, which were previously hidden by the Nb capping layer. The black curve corresponds to the X-ray's linear attenuation coefficient $\mu$ as derived from spectroscopic measurements published by Valencia and coworkers~\cite{Valencia2006_Medge_XMCD_MagnetoOpticalConstantsOfFeNiCo}. }.\label{fig_XAS_Medge}
\end{figure}

Initially, our XAS measurements around the Co M$_{2,3}$ edges (expected around \SI{60}{\electronvolt}~\cite{Thompson2001}) did not feature any peak, but instead a rather constant trend which is illustrated in Fig.~\ref{fig_XAS_Medge}.(a). This came as a surprise considering the good signal-over-noise ratio at the L$_3$ edge. Since the same observation was made on several \cogd beads, notably some for which L$_3$ edge XAS had been acquired, the only explanation for the aspect of our low-energy XAS is that the Co M$_{2,3}$ edges are hidden from the X-ray point of view, not the point of view of photoemission. In this respect, we find that a first hint of our Nb layer's influence is the ratio of the PEEM intensities recorded (on the bead) at the absorption peak and at the edge's foot. Indeed, it has been shown for PEEM-recorded XAS on 3D objects that this ratio is usually in excess of 1.3 for 3d ferromagnetic metals~\cite{Jamet2015,wartelle_phdthesis}, reaching values of 2 to 3 at the Co L$_3$ even in samples where Co is a minority element~\cite{wartelle_phdthesis}. In our \cogd beads, where we expect Co to be predominant, this intensity ratio is never higher than 1.5, which would be consistent with a capping thickness large enough to suppress photoemission from the underlying \cogd alloy~\footnote{We do not consider the (weak) X-ray absorption by the capping before photoemission in \cogd occurs, yet this decrease of intensity impinging on the magnetic alloy would only further suppress the photoemission.}. In other words: the L$_3$ edge data already suggests a non-negligible masking by the Nb capping layer.

At this stage, we point out that Nb's attenuation length below \SI{65}{\electronvolt} is below \SI{40}{\nano\meter}~\cite{Henke1993}. By contrast, it is in excess of \SI{130}{\nano\meter} around the Co L$_3$ edge. Also, it must be kept in mind that for the top, close to horizontal surface of the bead, the path length of X-rays through the capping is the latter's thickness enhanced by $1/\sin(\Psi)\simeq 3.6$ because of the beam's incidence angle $\Psi=$\ang{16}. This means that effective Nb thickness is on the order of \SI{10}{\nano\meter} or more, rather than the few nominally deposited nanometres. Concerning that thickness, we noted that the total Nb thickness  estimated to be etched by our Ar plasma sputtering is \SI{15}{\nano\meter}. Even if we assume the presence of a harder-to-etch, \emph{ca.} \SI{1.6}{\nano\meter} oxide layer~\cite{Marks1982_NativeSurfaceOxideNb} with a 4:1 ratio of sputtering yields in our conditions, we are led to consider a total NbO$_x$/Nb capping thickness of about \SI{10}{\nano\meter}.

%%%%%%%%%%%%%%%%%%%%%%%%%%%%%%%%%%%%%%%%%%%%%%%%%%%%%%%%%%%%%%%%%%%%%%%%%%%%%%%%
%%%%%%%%%%%%%%%%%%%%%%%%%%%%%%%%%%%%%%%%%%%%%%%%%%%%%%%%%%%%%%%%%%%%%%%%%%%%%%%%
%\textcolor{red}{To be detailed, I have only a few notes in the paper logbook}
%%%%%%%%%%%%%%%%%%%%%%%%%%%%%%%%%%%%%%%%%%%%%%%%%%%%%%%%%%%%%%%%%%%%%%%%%%%%%%%%
%%%%%%%%%%%%%%%%%%%%%%%%%%%%%%%%%%%%%%%%%%%%%%%%%%%%%%%%%%%%%%%%%%%%%%%%%%%%%%%%

Furthermore, in a first approximation, the XAS jump associated to an absorption edge is expected to scale with $\mu_\mathrm{on}-\mu_\mathrm{below}$ where $\mu_\mathrm{on}$ and $\mu_\mathrm{below}$ are the X-ray linear attenuation coefficients at the absorption peak and at its foot, respectively. If we compare the M$_{2,3}$ and L$_3$ edges of Co, we find that $(\mu_\mathrm{on}-\mu_\mathrm{below})/\mu_\mathrm{below}\sim 1.6$ at the M$_{2,3}$ edges~\cite{Valencia2006_Medge_XMCD_MagnetoOpticalConstantsOfFeNiCo}, whereas $(\mu_\mathrm{on}-\mu_\mathrm{below})/\mu_\mathrm{below}\sim 23$ at the L$_3$ edge~\cite{Nakajima1999}. 

Therefore, we consider that the combination of weaker (in a relative sense) jump in edge absorption and much stronger X-ray attenuation by the Nb capping are responsible for initially masking the Co M$_{2,3}$ edges. A crude estimate of the combination of both effects suggests a suppression of the edge jump's strength by a factor $\sim 33$, which is very likely to bring the Co M$_{2,3}$ edges below the noise level.

%To go into more details, we recall that Nb forms a surface oxide layer of \emph{ca.} \SI{1.6}{\nano\meter} according to Marks \emph{et al.} \cite{Marks1982_NativeSurfaceOxideNb}.

%%%%%%%%%%%%%%%%%%%%%%%%%%%%%%%%%%%%%%%%%%%%%%
%%%%%%%%%%%%%%%%%%%%%%%%%%%%%%%%%%%%%%%%%%%%%%
%Handbook of Chemistry and Physics
%Nb has a density of 8.57 g/cm^3
%NbO has a density of 7.3 g/cm^3
%NbO2 has a density of 5.9 g/cm^3
%Nb2O5 has a density of 4.6 g/cm^3
%%%%%%%%%%%%%%%%%%%%%%%%%%%%%%%%%%%%%%%%%%%%%%
%%%%%%%%%%%%%%%%%%%%%%%%%%%%%%%%%%%%%%%%%%%%%%
%%%%%%%%%%%%%%%%%%%%%%%%%%%%%%%%%%%%%%%%%%%%%%

%\textcolor{red}{NB: The , and due to the incidence angle, the $\sim$\SI{10}{\nano\meter} capping are equivalent to $\sim$\SI{36}{\nano\meter}}

\subsection{X-ray magnetic imaging}

\begin{figure}[!h]
\centering\includegraphics[width=\columnwidth]{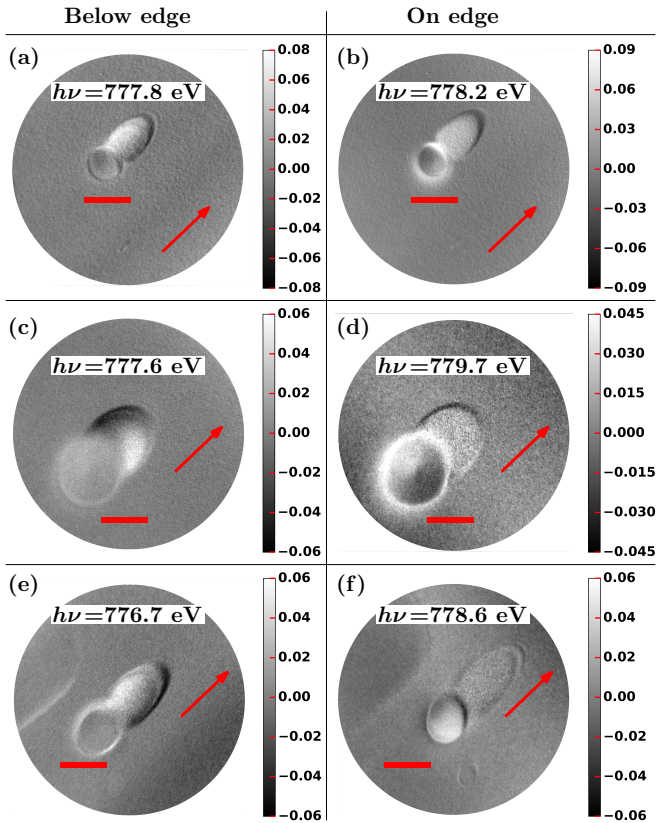}
\caption{Examplary XMCD-PEEM images acquired close to the Co L$_3$ edge, each row pertaining to one particular \cogd bead; (e-f) correspond to the one shown in Fig. \ref{fig_XAS_PEEM_bead}. In the left column, the photon energies are below (as far as \SI{1.9}{\electronvolt}) the absorption peak, and tuned in order to favour shadow XMCD-PEEM contrast. In the right column, the photon energies are at or slightly below the absorption peak so as to obtain XMCD-PEEM contrast from the top of the beads. For all images, the colour scales correspond to the plotted asymmetry ratio~$\mathcal{A}$. The arrows indicate the direction of the X-ray beam, and the scale bars correspond to \SI{1}{\micro\meter}. It must be noted that the focal plane was set to the substrate for shadow XMCD-PEEM and to the top of the beads for the ``on-edge'' measurements.}\label{fig_XMCD_PEEM_L3}
\end{figure}

Let us first review a few XMCD-PEEM images acquired at the Co L$_3$ edge. An overview of what we could observe at such photon energies is shown in Fig.~\ref{fig_XMCD_PEEM_L3}. When imaged far enough below the absorption peak, and with the microscope focused on the substrate, the \cogd beads typically yield bipolar shadow XMCD-PEEM patterns, as can be seen in the left column of Fig.~\ref{fig_XMCD_PEEM_L3}. Then, regular XMCD-PEEM imaging on the top surface of the beads performed with the photon energy at the maximum of absorption confirms the change of contrast from one side of the objects to the other, with the contrast inversions described by Jamet \emph{et al.}~\cite{Jamet2015}. The several occurrences of these bipolar patterns allow to confirm the preliminary results from MFM: vortex or at least vortex-like states seem to be predominant in our samples.

%%%%%%%%%%%%%%%%%%%%%%%%%%%%%%%%
%Fig preparation: mask from Fig_Sum_2022_06_04_022.pdf must be scaled by 0.815 to match the images in
% ./Figs_XMCD_PEEM_L3/Fig_XMCD_PEEM_L3_new_v3.pdf
%%%%%%%%%%%%%%%%%%%%%%%%%%%%%%%%

Close to the absorption peak, the transmission XMCD-PEEM contrast is lost in the bulk of the shadow because of the sample thickness instead of saturating~\cite{Jamet2015}, possibly because of harmonics contamination of the primary beam~\footnote{Even if only 1\% of the incident photons have twice the nominal photon energy $h\nu$, the X-ray attenuation length of our \cogd alloy is more than ten times larger at $2h\nu$ than at the Co L$_3$ edge. After several hundreds of nanometres, the number of photons in the harmonics beam exceeds that of the primary beam by many orders of magnitude.}. Thus, the shadow contrast in Fig.~\ref{fig_XMCD_PEEM_L3}.(b,f) is not reliable; the notable exception in Fig.~\ref{fig_XMCD_PEEM_L3}.(d) (with the correct contrast inversion between shadow and bead) is likely due to the slightly decreased absorption above the L$_3$ edge's white line peak.

However, we do see modulations in the shadow XMCD-PEEM contrast in Fig.~\ref{fig_XMCD_PEEM_L3}.(a,c,e). Such modulations \emph{inside the shadow} cannot result from errors in image alignment, unless (i) some variations in transmitted intensity occur near the shadow rim  for the three beads, \emph{and} (ii) alignment errors are every time along the same direction (roughly perpendicular to the X-ray beam). The combination of these two conditions seems unlikely. On the other hand, the modulation's asymmetry suggests a link to the asymmetry in average magnetization on both sides of the objects. Moreover, it must be kept in mind that for the images in Fig.~\ref{fig_XMCD_PEEM_L3}.(a,c,e), the X-ray absorption is small enough to allow significant signal-over-noise ratio for transmission XMCD-PEEM over the whole shadow, contrary to the cases described by Jamet \emph{et al.}~\cite{Jamet2015}. As a result, we can claim that these modulations are not artefacts related to saturation in absorption. Our conclusion is that these images are evidence for small-angle XRMS modulations on top of the usual ``geometrical-optics'' shadow XMCD-PEEM; in the following, we will refer to this effect as XRMS-PEEM.

Since the visibility of the fringes is rather low and their pattern is strongly confined near the shadow's rim, we then tune the photon energy to the Co M$_{2,3}$ edges. As was explained in Sec.~\ref{sec_Methods_PEEM_imaging} and Sec.~\ref{sec_Results_Spectro}, Co-sensitive image only became possible after significant sample sputtering, and with additional hindrances pertaining to the increased electron current in the PEEM's imaging column. This has an additional consequence on magnetic imaging, namely: the poorer spatial resolution impairs our accuracy in image alignment because all sharp features on the PEEM images (sample edges, substrate roughness \emph{etc.}) are blurred. In practice, we find a poorer reproducibility than for L-edge XMCD-PEEM; however, certain repeated images are in good agreement with one another.

\begin{figure}[!h]
\includegraphics[width=\columnwidth]{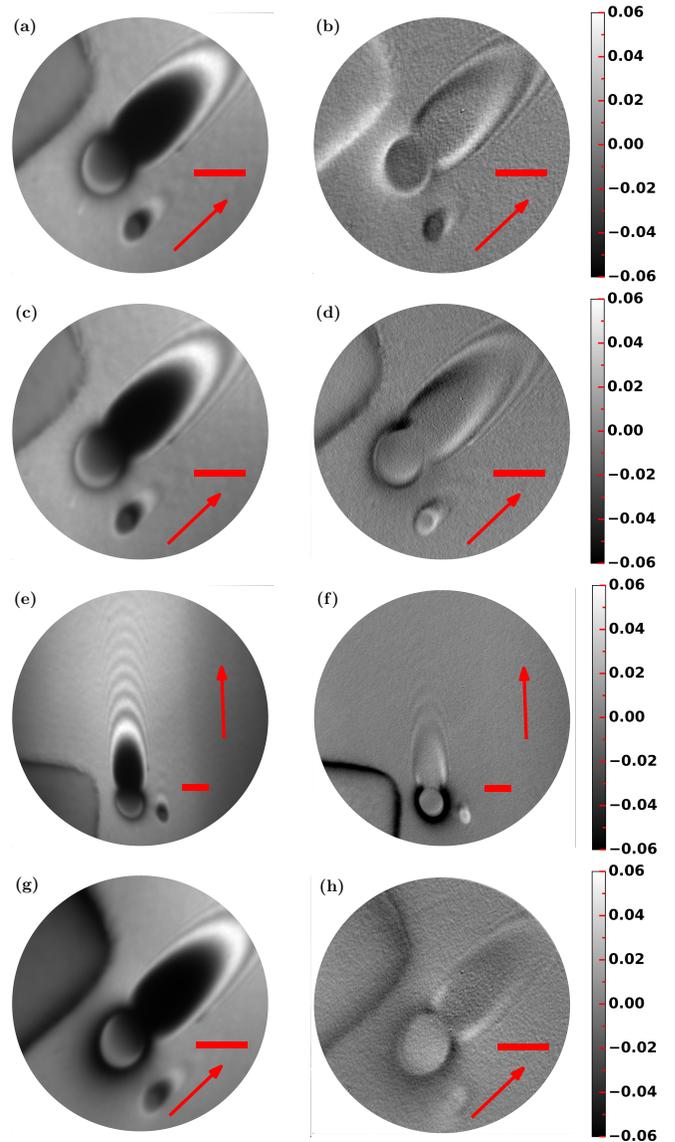}
\caption{Regular PEEM images without magnetic contrast (left column) and their corresponding magnetic images with the same colour range for $\mathcal{A}$ (right column). All these images pertain to the same bead as in Fig.~\ref{fig_XAS_PEEM_bead}.(b) Fig.~\ref{fig_XMCD_PEEM_L3}.(e-f), and have been acquired at the Co M$_{2,3}$ edge, the three first rows at \SI{60}{\electronvolt} and the last row (g-h) at \SI{65}{\electronvolt}. These are independent measurements separated by several other acquisitions. While a different magnification was used in (e-f), all scale bars are \SI{1}{\micro\meter}. The red arrows indicate the direction of X-ray propagation.}\label{fig_Medge_XRMS_PEEM}
\end{figure}

To illustrate first the stark differences in PEEM imaging, we present in Fig.~\ref{fig_Medge_XRMS_PEEM}.(a,c,e) several images acquired at a photon energy of \SI{60}{\electronvolt} of the same bead as before [the one shown in Figs.~\ref{fig_XAS_PEEM_bead}.(b) and \ref{fig_XMCD_PEEM_L3}.(e-f)]. It can be noted for instance from  Fig.~\ref{fig_Medge_XRMS_PEEM}(a) that the immediate vicinity of the bead appears distorted: there is a radial pattern around its perimeter that does not match the substrate roughness elsewhere. This enhanced optical aberration due to a 3D object is attributed to the instrumental complications described in Sec.~\ref{sec_Methods_PEEM_imaging}. Beyond this, one can clearly see a much clearer and more extended interference pattern behind the bead, notably in Fig.~\ref{fig_Medge_XRMS_PEEM}(e), where the magnification is lower than in Figs.~\ref{fig_Medge_XRMS_PEEM}(a,c).

At this stage, let us examine the images  that display the corresponding asymmetry ratio $\mathcal{A}$ in Figs.~\ref{fig_Medge_XRMS_PEEM}(b,d,f). While there are minute differences between them, the antisymmetry in the fringe pattern (perpendicular to the direction of X-ray propagation) is the same across the three images, and it is plausible considering the previous magnetic images of that \cogd bead at the L$_3$ edge. 

In order to support our hypothesis that this contrast is of magnetic origin, we present an additional pair of images of the same bead in Figs.~\ref{fig_Medge_XRMS_PEEM}(g,h), this time at a photon energy of \SI{65}{\electronvolt} \emph{i.e.} at a significant offset with respect to the previous images, as can be surmised from Fig.~\ref{fig_XAS_Medge}.(b). One would expect quite a strong reduction in XRMS strength upon moving the photon energy by an amount close to the edge jump's width, and the fringe contrast in Fig.~\ref{fig_Medge_XRMS_PEEM}(h) is clearly weaker than those in Figs.~\ref{fig_Medge_XRMS_PEEM}(b,d,f).

%\begin{figure}[!h]
%\includegraphics[width=0.4\textwidth]{2022_05_30_015_xas_bead_shadow.png} \hfill \includegraphics[width=0.4\textwidth]{2022_06_04_033_xas_2023_02_06_11h02mn43s_bead_shadow_warrows.png}
%\caption{XAS traces around the Co M$_{2,3}$ edges obtained on distinct CoGd beads before (left) and after (right) several rounds of Ar plasma sputtering. It must be noted that several spectra qualitatively similar to the left one were obtained on different beads. The significant (in terms of relative change) upturn in PEEM intensity starting a bit before \SI{60}{\electronvolt} is attributed to the Co M$_{2,3}$, which was previously hidden by the Nb capping layer. NB: The Nb attenuation length at these photon energies is always (significantly) below \SI{50}{\nano\meter}, and due to the incidence angle, the $\sim$\SI{10}{\nano\meter} capping are equivalent to $\sim$\SI{36}{\nano\meter}.}\label{fig_XAS_Medge}
%\end{figure}

%\newpage

%\begin{figure}[!h]
%\centering\includegraphics[width=0.85\textwidth]{Fig_normedBeadXAS_MuFromValenciaCoMedgeBeta.png}
%\caption{Measured M$_{2,3}$-edge XAS from Fig. \ref{fig_XAS_Medge} after several roudns of sputtering, plotted with the linear absorption coefficient of Co as obtained based on the refractive index data measured by Valencia \emph{et al.} \cite{Valencia2006_Medge_XMCD_MagnetoOpticalConstantsOfFeNiCo}. Some differences in the post-peak behaviour might originate from \emph{e.g.} the dispersion of absorption in Gd (in the bead) or in Si (on which the normalizing background is measured).}
%\end{figure}

\begin{figure}[h!]
\includegraphics[width=\columnwidth]{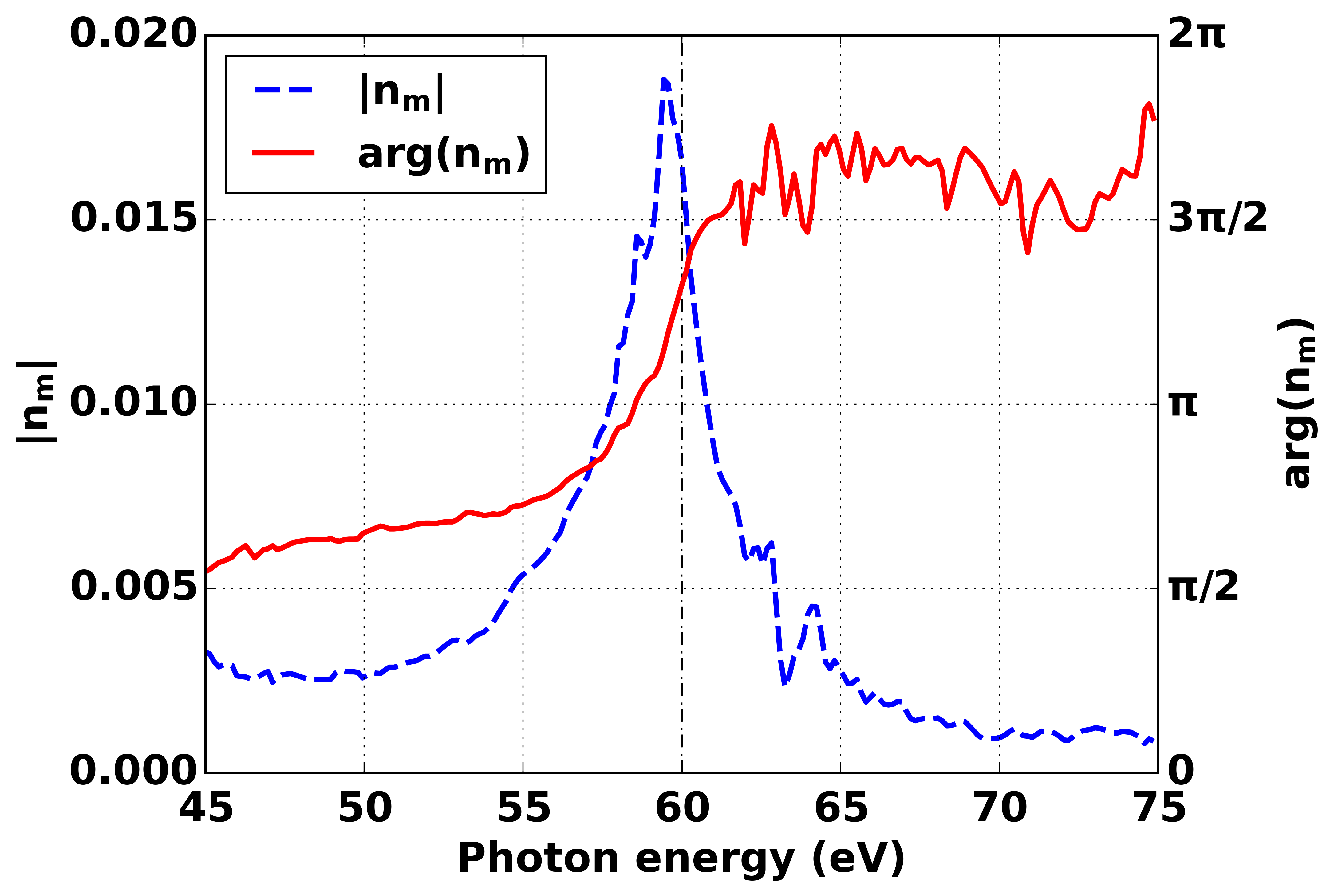}
\caption{Norm ($|n_m|$, blue dashed curve) and complex argument [$\mathrm{arg}(n_m)$] of the magnetic contribution to Co's refractive index near the M$_{2,3}$ edges, computed from the results reported by Valencia \emph{et al.}~\cite{Valencia2006_Medge_XMCD_MagnetoOpticalConstantsOfFeNiCo}.}\label{fig_nm_Co_Medge}
\end{figure}

Using the magnetic contributions $\Delta\delta$ and $\Delta\beta$ to the X-ray refractive index of Co around its M$_{2,3}$ edges determined by Valencia \emph{et al.}~\cite{Valencia2006_Medge_XMCD_MagnetoOpticalConstantsOfFeNiCo}, we compute the norm $|n_m|$ and complex argument $\mathrm{arg}(n_m)$ of $n_m=-\Delta\delta+i\cdot\Delta\beta$ and plot them in Fig.~\ref{fig_nm_Co_Medge}. As can be seen on this graph, the strength of XRMS (behaving as $|n_m|$) is close to maximum at \SI{60}{\electronvolt}, whereas it is about six times lower at \SI{65}{\electronvolt}. Taking this into account, we can now understand the contrasts in Figs.~\ref{fig_Medge_XRMS_PEEM}.(b,d,f,h) as follows. The quasi-bipolar pattern in Figs.~\ref{fig_Medge_XRMS_PEEM}.(b,d,f) at \SI{60}{\electronvolt} originates at least predominantly from M-edge XRMS, with part of the differences between those images caused by slight errors in image alignment, while the significantly weaker pattern in Figs.~\ref{fig_Medge_XRMS_PEEM}.(h) is a combination of low XRMS and imperfect drift correction. We thus conclude that the few-percents contrast which has opposite signs on opposite sides of the \cogd bead in Figs.~\ref{fig_Medge_XRMS_PEEM}.(b,d,f) is XRMS-PEEM. 

\subsection{Analytical and numerical modelling}

\begin{figure}[h!]
\centering\includegraphics[width=\columnwidth]{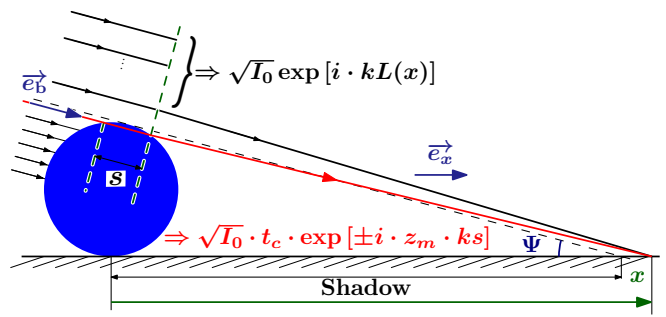}
\caption{Schematics for the two-source analytical model. The \cogd bead is represented in blue, and the relevant typical length $s$ traversed by X-rays through it is shown between the yellow dashed lines. The black dashed line indicates the trajectory of X-rays at the geometrical-optics rim of the beads' shadow.}\label{fig_analytical_model}
\end{figure}

In the following, we will explore a very simplistic model for the interference pattern observed in PEEM beyond the shadow of the \cogd beads, at the Co M$_{2,3}$ edges. Our objective is merely to obtain physical insight into the relation between the XRMS-PEEM fringe patterns and the Fresnel diffraction from the sample. To that end, our coarse approach consists in approximating the complex wave propagation and scattering from a 3D object of unknown shape with a two-source interference problem. A schematic view of the situation of interest is presented in Fig.~\ref{fig_analytical_model}, with a spherical \cogd bead (in blue) on a substrate and an X-ray beam impinging on it with the experimental incidence angle $\Psi$.

Let us now describe our model's features in more detail. First of all, we omit any three-dimensional aspect of the problem. Then, considering how opaque the scatterer is (with attenuation lengths of at most $\simeq$\SI{12}{\nano\meter}), we neglect X-ray transmission for most of the bead and only keep contributions from parts with a low path length $s$ traversed through the alloy. These contributions are gathered and considered as the first of our two sources.

The second source is of course the part of the beam that (in a geometrical-optics sense) does not impinge on the bead. For illustration purposes, a ray from the first source is shown in Fig.~\ref{fig_analytical_model} (in red) intersecting a ray from the second one at a point on the substrate with abscissa $x$, behind the rim of the geometrical-optics shadow of the bead (in blue). As we are not aiming at a quantitative expression for the X-ray intensity as a function of abscissa $x$ on the substrate, we will instead leave as unknown the path difference $L(x)$ between the first source and that part of the beam which passes above it from here onwards referred to as the second source. In doing so, we absorb the integration over that second part of the incident wave into a beam amplitude $\sqrt{I_0}$ and the path difference~\footnote{We neglect necessary variations in amplitudes as these would only add the complexity of a spatially varying oscillation envelope with no further insight.} $L(x)$. With respect to $\sqrt{I_0}$, we will use the modulus of a propagation prefactor $t_c$ for the first source's amplitude so as to take into account the sample's non-magnetic refractive index (we restrict ourselves to small-angle scattering) $n_c=1-\delta+i\cdot\beta$.

From this point on, we must determine the expression for the Fresnel diffraction that is caused only when charge scattering is considered. As a second step, we shall ``turn on'' XRMS by considering the magnetic contributions to our sample's refractive index, and recalculate the expression for the X-ray intensity on the substrate. Our goal is to assess whether XRMS  modulates rather the oscillations' amplitude, resulting in a magnetic asymmetry $\mathcal{A}$, or phase-shifts the patterns associated to each polarization depending on magnetization.

Let us denote as before $n_m=-\Delta\delta+i\cdot\Delta\beta$, $m_\mathrm{b}=s^{-1}\int_{0}^s \overrightarrow{m}\cdot \overrightarrow{e_\mathrm{b}}\mathrm{d}u$ the average projection of magnetization along the beam with propagation direction $e_\mathrm{b}$, and $z_m=n_m\cdot m_\mathrm{b}$. For the sake of simplicity, we will consider that magnetization can only yield either $+m_\mathrm{b}$ or $-m_\mathrm{b}$ as projections. We may now write the amplitudes of our two sources $a_{\pm}$ and $a_0$ as:

\begin{equation}\label{eq_Model_amplitudes}
\begin{split}
a_{\pm}	&=\sqrt{I_0}\cdot t_c \cdot\exp{[\pm i\cdot z_m\cdot ks]}, \\
a_0		&=\sqrt{I_0}\cdot\exp{\Big[i\cdot kL(x)\Big]}.
\end{split}
\end{equation}

With Eq.~\eqref{eq_Model_amplitudes} in mind, we can write down the X-ray intensities on the substrate $I_\pm$ as:

\begin{equation}
\begin{split}
I_\pm(x)	&=	|a_0+a_\pm|^2 \\
			&=	I_0\cdot\Big[1+2\cdot\mathrm{Re}\Big(t_\mathrm{c}\cdot e^{\pm i\cdot z_m ks} e^{-i\cdot kL(x)}\Big) \\
			&	\quad +|t_\mathrm{c}|^2 e^{\pm 2 \mathrm{Im}(z_m)\cdot ks}\Big].
\end{split}
\end{equation}

If we use the values of $\Delta\delta$ and $\Delta\beta$ from Valencia \emph{et al.}~\cite{Valencia2006_Medge_XMCD_MagnetoOpticalConstantsOfFeNiCo}, we have  $\Delta\delta= 8.9\cdot 10^{-3}$ and $\Delta\beta=-1.4\cdot 10^{-2}$ for a photon energy $h\nu=$\SI{60}{\electronvolt}; by contrast,  $\delta=-2.0\cdot10^{-2}$ and $\beta=1.6\cdot10^{-1}$. Since the attenuation length is lower than \SI{12}{\nano\meter}, we may consider values of $s$ of, at most, \SI{24}{\nano\meter}. In that case, $|t_c|\simeq 0.31$. With a wavelength $\lambda=$\SI{20.6}{\nano\meter}, we find $n_m \cdot ks \simeq  -0.065-i\cdot0.10$. Since $|m_\mathrm{b}|\leq 1$ and $|n_m|\cdot ks=0.12$, we may approximate $\exp{[\pm i\cdot z_m ks]}\simeq 1 \pm i\cdot z_m ks$.

%For s 15 nm,$n_m \cdot ks \simeq  -0.041-i\cdot0.064$, abs 0.076

After a bit of algebra, we retrieve the expressions for the magnetic and non-magnetic fringe patterns. For the sake of normalization but also of computing an asymmetry ratio as we do in experiments, we use $[I_+(x)-I_-(x)]/(2I_0)$ for the former, and $[I_+(x)+I_-(x)]/(2I_0)$ for the latter. To first order in $|t_\mathrm{c}|$, these quantities read:

\begin{equation}
\begin{split}
\frac{I_+(x)-I_-(x)}{2I_0}=	&2m_\mathrm{b}\cdot ks\cdot\mathrm{Re}(t_\mathrm{c}\cdot n_m)\cdot \cos{\Big[kL(x)\Big]} \\
							&+2m_\mathrm{b}\cdot ks\cdot\mathrm{Im}(t_\mathrm{c}\cdot n_m)\cdot \sin{\Big[kL(x)\Big]}, \\
\end{split}
\end{equation} \\ and:

\begin{equation}
\begin{split}
\frac{I_+(x)+I_-(x)}{2I_0}=	&2\mathrm{Re}(t_\mathrm{c})\cdot \cos{\Big[kL(x)\Big]} \\
							&+2\mathrm{Im}(t_\mathrm{c})\cdot \sin{\Big[kL(x)\Big]}.
\end{split}
\end{equation}

Denoting $\varphi_t=\mathrm{arg}(t_{\mathrm{c}})=-ks\cdot \delta$ and $\varphi_m=\mathrm{arg}(n_m)$, we may rewrite the above equations as:

\begin{equation}
\frac{I_+(x)-I_-(x)}{2I_0}=	2m_\mathrm{b}\cdot ks\cdot |t_\mathrm{c}|\cdot |n_m| \cdot \cos{\Big[kL(x)-\varphi_t-\varphi_m\Big]}
\end{equation} \\ and:

\begin{equation}
\frac{I_+(x)+I_-(x)}{2I_0}=2 |t_\mathrm{c}|\cdot \cos{\Big[kL(x)-\varphi_t\Big]}.
\end{equation}

With this, our model indicates that XRMS leads to intensity modulations $\propto |n_m|\cdot m_\mathrm{b}$ as expected, and phase-shifted with respect to the non-magnetic fringe pattern by $-\varphi_m$. At \SI{60}{\electronvolt}, $-\varphi_m\simeq$-\ang{240}, suggesting that we might expect close to a phase quadrature experimentally. A very accurate match would however require a very careful calibration of the X-ray beam's energy beforehand since $\varphi_m$ varies by $\pi/2$ over \emph{ca.} \SI{3}{\electronvolt} according to Fig.~\ref{fig_nm_Co_Medge}. As an illustration, we present in the Supplemental Material a comparison between the values of $\Delta\delta$ and $\Delta\beta$ obtained by Valencia \emph{et al.}~\cite{Valencia2006_Medge_XMCD_MagnetoOpticalConstantsOfFeNiCo}, and those reported by Willems \emph{et al.} \cite{Willems2019_XrayRefractiveIndexOfCo_Ni_Fe_atMedges}. 

It must be noted that the estimated magnetic phase shift does not depend on the precise values of $s$ or $m_\mathrm{b}$ (only the overall amplitude does), which is comforting.  To conclude the discussion of our coarse analytical model, we propose the following interpretation: due to the larger contribution of XMCD  with respect to XMCB ($|\Delta\beta|\simeq 1.6\cdot\Delta\delta$), the largest magnetic contrast are obtained close to when the non-magnetic fringe patterns reaches its average value (for $kL(x)-\varphi_t \equiv \pi/2\quad [\pi]$ ) because then the interference term is most sensitive to variation in amplitudes from the first source.

\begin{figure}[h!]
\centering\includegraphics[width=\columnwidth]{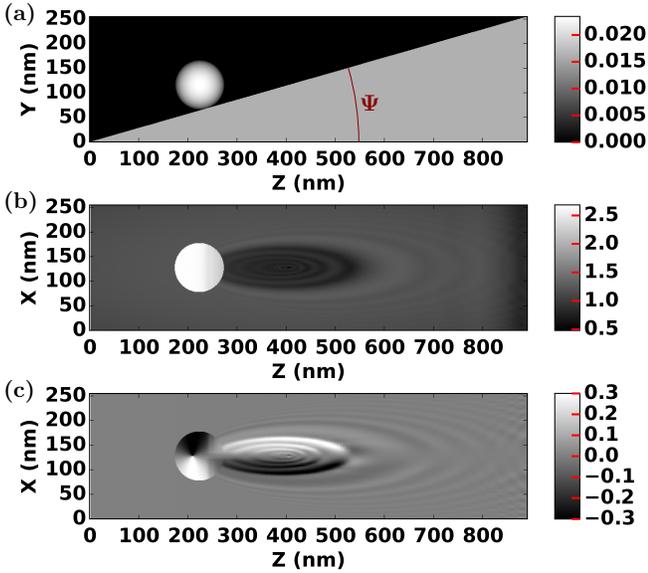}
\caption{(a) View from the side of the simulated sample, in terms of averaged linear attenuation coefficient. (b) computed PEEM-like image considering charge scattering only; a logarithmic scale is used for the sake of fringe visibility in the shadow. (b) XRMS pattern computed as normalized difference between PEEM-like images for opposite polarisations.}\label{fig_numerical_model}
\end{figure}

In order to look deeper into XRMS-PEEM, we now tackle the problem with a numerical model. Considering a rectangular simulation box containing a perfectly spherical magnetic bead on a substrate and with vacuum elsewhere, we compute the propagation of X-rays in an angular spectrum approach~\cite{Schafer1989_AngularSpectrumApproach}, with the incidence angle $\Psi$ as in experiments. The diameter choice is a compromise between object size (as the angular spectrum approach would become computationally too heavy for micron-sized beads) and extent of the scattering pattern behind the bead shadow.  Then, we obtain the expected PEEM image based on the absorption coefficient on the surface of the bead or on the substrate. In practice, we compute the X-ray propagation along the complete long axis of our simulation box, while the substrate is tilted by an angle $\Psi$, as illustrated in Fig.~\ref{fig_numerical_model}. We then obtain an image similar to experimental PEEM data based on the wave amplitudes on the topmost surfaces, weighted by the local absorption coefficient. As far as magnetism is concerned, the bead is made to host a simple configuration, with an analytical vortex ansatz whose core points perpendicular to the substrate.

In terms of X-ray physics, the relevant refractive indices for the substrate (Si) and for the magnetic material (considered to be Co$_{80}$Gd$_{20}$) are computed based on the tabulated values by Henke \emph{et al.}~\cite{Henke1993} as well as resonant magnetic scattering factors from the Dyna software package~\cite{Elzo2012_Dyna_ReferencePaper}, and finally the magneto-optical constants of Co determined by Valencia and coworkers. We first present results for the Co L$_3$ edge, with a bead diameter of \SI{100}{\nano\meter}, at a photon energy which maximizes XRMS (in the $|n_m|$ sense).

\begin{figure}[h!]
\centering\includegraphics[width=\columnwidth]{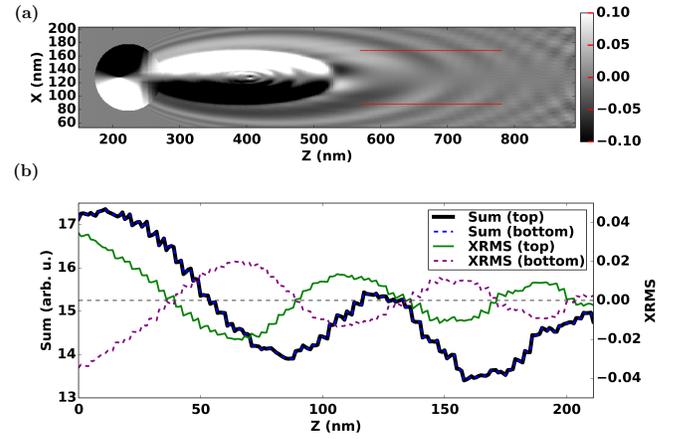}
\caption{(a) Zoom-in on Fig.~\ref{fig_numerical_model}.(c); the red lines indicate the profiles taken across this image. (b) Green dashed and purple full curves: profiles extracted from the lines drawn in (a). Full black and dashed blue curves: corresponding profiles extracted from the non-magnetic image.}\label{fig_numerical_model_profiles}
\end{figure}

 First of all, as can be seen on Fig.~\ref{fig_numerical_model}.(b), we do retrieve fringes mostly outside of the shadow; the ones inside are made visible by the chosen logarithmic scale. We point out that the finite size of our box along the height (Y direction) is one of the main reasons for the artefacts present on the right-hand side of the simulated PEEM-like images. Then, the computed asymmetry ratio $\mathcal{A}$ in Fig.~\ref{fig_numerical_model}.(c) shows XRMS contrast both inside and outside the bead's shadow, as expected. The expected shadow XMCD-PEEM contrast is modulated, and beyond the shadow's rim, we see an asymmetric fringe pattern reminiscent of our experimental images. The latter can be investigated further with suitable line profiles along the beam direction; we therefore proceed in Fig.~\ref{fig_numerical_model_profiles}.(a), where two such profiles are drawn on the simulated XRMS image as red lines. They have been placed symmetrically around the middle of the simulation box. Both have been extracted on the XRMS image and on the non-magnetic image, and the four corresponding traces are plotted in Fig.~\ref{fig_numerical_model_profiles}.(b). As must be the case, the magnetic traces are opposite. Moreover, we can observe a significant phase shift between the latter and the profiles on the non-magnetic image. However, it is not found constant across the simulated oscillations; typical estimates are \SIrange{60}{80}{\degree}.

\begin{figure}[h!]
\centering\includegraphics[width=\columnwidth]{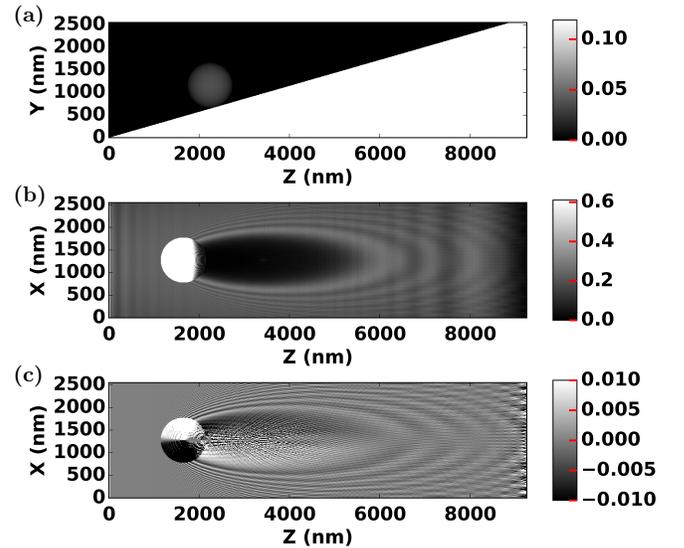}
\caption{(a) View from the side of the simulated sample at \SI{60}{\electronvolt}, in terms of averaged linear attenuation coefficient. (b) computed PEEM-like image considering charge scattering only; here the intensity scale is linear. (b) XRMS pattern computed as normalized difference between PEEM-like images for opposite polarisations.}\label{fig_numerical_model_Medge}
\end{figure}

At this stage, we perform similar simulations at a photon energy of \SI{60}{\electronvolt} in a \SI{1}{\micro\meter}-diameter bead. The vortex ansatz it hosts has been rescaled accordingly without further modifications. The results are shown in Fig.~\ref{fig_numerical_model_Medge} similarly to the previous ones. An important conclusion from the simulated XRMS-PEEM image is the change of sign of contrast for the fringes outside the shadow compared to the L$_3$ edge simulations. This stems directly from the opposite signs of $\Delta\beta$ \cite{Elzo2012_Dyna_ReferencePaper,Valencia2006_Medge_XMCD_MagnetoOpticalConstantsOfFeNiCo}, and is in agreement with the change observed on the bead from Fig.~\ref{fig_XMCD_PEEM_L3}.(e-f) when moving to the Co M$_{2,3}$ edges.

\section{Conclusion}

In the course of this work, we have performed magnetic imaging using X-ray PEEM and observed small-angle X-ray Resonant Magnetic Scattering at the Co M$_{2,3}$ edges. These experiments have been carried out on three-dimensional objects, namely \cogd beads, which we have taken care to characterize beforehand, including with more conventional L$_3$-edge (shadow) XMCD-PEEM. Combined with the pristine metallic character of Co revealed by XAS, this first imaging scheme has confirmed the suitability of this sample for an investigation of M-edge XRMS. Indeed, the vortex-like configurations that have been evidenced both from direct XMCD-PEEM views and from the telltale shadow contrast ensure that the incoming X-rays probe opposite projections of magnetization. Our expectation was therefore to find an asymmetry in the pronounced Fresnel diffraction at the Co M$_{2,3}$ edges. After careful image analysis, a plausible contrast has been extracted; its magnetic origin is supported not only by its decrease above the absorption edge but also by our modelling. Coarse as the latter may be, it does hint at the same asymmetry as in our experiments. Unfortunately, our numerical simulations indicate that an accurate measurement and interpretation of the XRMS pattern's phase shift with respect to the non-magnetic fringe pattern requires more than the simple analytical model we presented. On the other hand, our L$_3$-edge shadow XMCD-PEEM results suggest that small-angle XRMS can be sizeable and affect non-negligible parts of the shadow area~\cite{Jamet2015}, thus compromising efforts towards \emph{e.g.} shadow XMCD-PEEM-based vector magnetic tomography using laminography~\cite{Donnelly2019_ReviewOn3DxRayMagImaging_Tomography}.

However, it must be kept in mind that the photon-in, electron-out scheme is not the only one available in an X-ray PEEM. Indeed, the reflection geometry and CDI phase retrieval approaches have recently been successfully demonstrated on artificial spin ice systems~\cite{Mentes2020_XRDinSPELEEMonArtifSpinIces}. It stands to reason that the difficulty in reaching sufficient oversampling may be drastically suppressed by utilizing XRMS at M edges compared with L edges of 3d magnetic transition metals. In this respect, our investigation shows that magnetic contrast at this energies can already be accessed in the Fresnel diffraction of 3D samples, despite instrumental challenges which are specific to the electron imaging. Therefore, a longer-term prospect of this study is M-edge CDI in a reflection geometry. In addition to the possibilities of imaging under field~\cite{Mentes2020_XRDinSPELEEMonArtifSpinIces}, a very efficient use of photons can be foreseen thanks to the increased X-ray coherence as well as the absence of resolution degradation at large flux~\cite{Locatelli2011_ChargingEffectsInPEEMdueToLargeFlux}. Finally, the strong absorption is expected to be beneficial in the study of (ultra-)thin films, for which X-ray PEEM provides a valuable sensitivity to in-plane magnetization.

\section{Acknowledgements}

We thank Andrea Locatelli, Tevfik Onur Mente\c{s} and Matteo Jugovac (ELETTRA) for assistance in the XMCD-PEEM measurements as well as fruitful discussions. We acknowledge the Agence Nationale de la Recherche for funding under project number ANR-19-CE42-0013-05.

%\newpage

%\bibliographystyle{unsrt}
\bibliography{XRMS_PEEM_CoGd.bib}
\end{document}